\documentclass[12pt]{iopart}
\usepackage{iopams}
\usepackage{setstack}
\usepackage{lmodern} 
\usepackage{graphicx}
\usepackage{epigraph}
\usepackage{amssymb}
\usepackage{paralist}
\usepackage{color}
\usepackage{xfrac} 

\usepackage{amscd}
\usepackage{dsfont}
\usepackage{bm}

\def\HI{H_{\scriptscriptstyle\mathrm{I}}}
\def\HK{H_{\scriptscriptstyle\mathrm{K}}}
\def\UI{U_{\scriptscriptstyle\mathrm{I}}}
\def\UK{U_{\scriptscriptstyle\mathrm{K}}}
\def\UIh{U_{\scriptscriptstyle\mathrm{I}}}
\def\UKh{U_{\scriptscriptstyle\mathrm{K}}}

\def\trace{\Tr}

\def\Im{ {\mathrm{Im}}}

\newif\ifdraft
\draftfalse

\DeclareGraphicsExtensions{.pdf, .png, .bmp, .jpg}

\newcommand{\iu}{\mathrm{i}}
\newcommand{\eu}{\mathrm{e}}

\newcommand{\eg}{{\em e.g.}\ }
\newcommand{\ie}{{\em i.e.}\ }

\newcommand{\teq}{\! = \!}
\newcommand{\hatN}[1]{\hat{#1}}

\newcommand{\binom}[2]{{#1 \choose #2}}
\newcommand{\operatorname}[1]{\text{#1}}
\newcommand{\eqref}[1]{(\ref{#1})}

\begin{document}
\title[Particle-Time Duality I]{Particle-Time Duality in the Kicked Ising Chain I: The Dual Operator}

\author{M. Akila,  D. Waltner, B. Gutkin, T. Guhr}
\address{
Faculty of Physics, University Duisburg-Essen, Lotharstr. 1, 47048 Duisburg, Germany}
\begin{abstract}
We demonstrate that the dynamics of kicked spin chains possess a remarkable duality property. The trace of the unitary evolution operator
for $N$ spins at time $T$ is related to one of a non-unitary evolution operator for $T$ spins at time $N$. We investigate the spectrum of this dual operator with a focus on the different parameter regimes (chaotic, regular) of the spin chain.
We present applications of this duality relation to spectral statistics in an accompanying paper.
\end{abstract}

Pacs: 03.65.Aa, 05.45.Mt, 05.45.Pq

\maketitle

\section{Introduction}
The connection between  quantum   and  classical dynamics of Hamiltonian systems is  revealed in the semiclassical limit $\hbar \to 0$,  where the  effective dimension of the Hilbert space 
grows indefinitely. For propagation up to times of the order of the Heisenberg time $T_H=2\pi \hbar/\Delta$, the classical-quantum correspondence has been extensively used to explore the spectral 
correlations on the scale of the mean level spacing $\Delta$, see \cite{haake,haake2,haake3,stoeckmann}. Particularly, for generic single particle systems with classically chaotic dynamics the universal spectral correlations were  
established via the statistical correlations of long periodic orbits. On the other hand, at very  short (classical) time scales the contribution to spectral correlation functions  are 
provided by  very few  periodic orbits, implying that universality cannot be expected \cite{Berry}.     This conclusion, however,   should be taken with a grain of salt  for systems  composed of a 
large number $N$  of  particles,
extensively studied in recent years \cite{regen1, haemmer, haemmer2, dubert,buchleit1,znidaric, buca}.
First, for many-particle systems 
 even periodic orbits with  short periods $T$ might acquire large actions due to the interactions between particles. Second,   in chaotic systems   periodic orbits  typically proliferate   
 exponentially with $N$, such that their  number becomes very large already  for short $T$. It is therefore a natural question whether  a kind of universality can be expected in the 
 regime of short times $T$ but large number of particles $N$.

To resolve this question we introduce a duality-based approach.
Its main idea can be understood on a classical level: consider a system of \(N\) particles  with positions \(x_n\) and momenta \(p_n\). At discrete points in time the system is kicked by a ``potential'' consisting of both harmonic nearest neighbor interaction between the particles and an arbitrary field, acting on each particle separately, \ie
\begin{equation}
V_n(x_n,x_{n+1},t)=\left(\frac{\omega^2}{2}(x_n-x_{n+1})^2+v(x_n)\right)\sum_{\tau=-\infty}^\infty \delta(t-\tau)
\end{equation}
for two neighboring particles. For the resulting Hamiltonian
\begin{eqnarray}
 H(\mathbf{x},\mathbf{p},t)=\sum_{n=1}^N \left(\frac{p_n^2}{2 m} +V_n(x_n,x_{n+1},t)\right)
\label{prototypemodel}
\end{eqnarray}
we also impose
cyclic boundary conditions which render \(H(\mathbf{x},\mathbf{p},t)\) invariant under the shift \(n\to n+1\), in particular we have \(x_{N+n}\teq x_n\).
This model defined by (\ref{prototypemodel}) can  be thought of as  $N$ interacting kicked rotors. 

Written in coordinate form the  dynamics of the model is described by the  map
\begin{equation}
 m\left[x_{t+1, n}+ x_{t-1,n}-2x_{t, n}\right] + \omega^2\left[ x_{t, n+1}+ x_{t,n-1}-2x_{t, n}\right]=v'(x_{t,n})\,,
\end{equation}
where $x_{t,n}$ stands here for the coordinate of the $n$-th particle at time $t$. 
Remarkably, these  equations
stay invariant under the exchange of the particle and time indices $n\leftrightarrow t$ if one simultaneously exchanges the model  parameters $m \leftrightarrow \omega$. In particular, this implies that  all $T$-periodic orbits of the $N$-particle system with parameters $(m,\omega)$  are in one-to-one correspondence with $N$-periodic orbits of a $T$-particle system with exchanged parameters $(\omega,m)$. Such a duality relation was first observed in the model of coupled cat maps \cite{Gutkin}. It was noticed there that such a 
property of the  classical Hamiltonian  must imply a relation between the corresponding quantum evolutions. Indeed,
since the trace of the quantum propagator  can be expressed in the semiclassical limit through periodic orbits of period \(T\), 
one might expect in that limit the relation
\begin{equation}
 \trace{U_N(T)} \sim
 \trace{\tilde{U}_T(N)} \label{qduality}
\end{equation}
between the propagator $U_N(T)$ of the $N$-particle system for the time $T$ and the propagator 
$\tilde{U}_T(N)$  for the dual  system of $T$ particles and time $N$. The importance of relation (\ref{qduality}) stems
from the fact that it allows one to study  the regime  of short times $T$ and a large number of particles $N$ by converting it into 
one  of large times and a small number of particles. The last one  is often amenable to treatment by standard semiclassical methods. 

Here, we explore the existence of the relation (\ref{qduality}) in a different class of models, namely kicked spin-1/2 chains first introduced in \cite{prosenJt-2}. In this class of models an underlying classical limit does not exist. Nevertheless, as we show later on,
an analogue of (\ref{qduality}) holds in this case as well.
We can exploit this to express the traces of the short-time propagator for large $N$ in terms of the (by modulus) largest eigenvalues of its dual counterpart.

The outline of the paper is as follows: in section 2 we introduce the evolution operator of the kicked spin chain and 
in section \ref{sec:dualityRel} its dual complement. We analyze the spectrum of this dual operator in different parameter regimes in section \ref{sec:specDOp} and conclude in section \ref{section5}. Technical details are relegated to the appendix. In an accompanying paper we will discuss applications of this trace duality to the calculation of spectral densities and form factors in the limit of large particle numbers.

\section{Kicked Spin Chain Model}

The KIC model \cite{prosenJt-2,prosen2,prosen2007,prosenB3-d}
is defined as  a ring of  $N$ spins with homogeneous nearest neighbor coupling and an additional magnetic field kicking the system. The Hamiltonian 
\begin{equation}
H=\HI+\HK \sum_{\tau=-\infty}^\infty\delta(t-\tau)
\end{equation}
is therefore a sum of two contributions. The first part,
\begin{equation}
 \HI=J\sum_{i=1}^N \hat{\sigma}^z_{i}\hat{\sigma}^z_{i+1}
\,,
\end{equation}
is the standard Ising-Hamiltonian
for spins \(\bm{\hat{\sigma}}_{i}\teq(\hat{\sigma}^x_{i}, \hat{\sigma}^y_{i},\hat{\sigma}^z_{i})\) where the components \(\hat{\sigma}^j_{i}\) are the Pauli matrices for spin \(i\). The parameter \(J\) governs the coupling strength.
The second part contains a homogeneous magnetic field \(\bm{b}=(b^x,b^y,b^z)\) providing the action
\begin{equation}
 \HK=\sum_{i=1}^N \bm{b} \cdot \bm{\sigma}_{i}
\end{equation}
of the kicks.
For convenience, we measure time in dimensionless units.
Without loss of generality the  magnetic field  can be restricted to the $(x,z)$ plane, such that 
\begin{equation}
\bm{b}= (b\sin\varphi, 0, b\cos\varphi)\,.
\end{equation}
The Floquet operator for one period of the time evolution is thus given by the product
\begin{equation}
U_N=\UIh(J) \UKh(b,\varphi)\,,
\end{equation}
where
\begin{eqnarray}
\UIh(J)=\exp(-\iu \HI)
\qquad \text{and}\\
\UKh(b,\varphi) = \exp(-\iu \HK)
\end{eqnarray}
correspond to free evolution and kicks, respectively.

Depending on the strength of the coupling $J$ and the magnetic field $\bm{b}$ the KIC  shows different  regimes ranging from integrable dynamics  to full chaos.  Given the lack of a classical limit the definition of chaoticity is based on the spectrum of \(U_N\), the system is said to be chaotic if the nearest neighbor spacing distributions obeys Wigner-Dyson statistics. In \cite{prosen2} a dynamical characterization based on correlators between spins was found to be consistent with the above approach.
Due to symmetries it is sufficient to consider the parameters \((J,\,b,\,\varphi)\) in the interval of \(0\) to \(\pi/2\) only.
Further on, the model is exactly integrable if either $b^x$ or $b^z$ vanishes, \ie $\varphi\teq 0, \pi/2$. 
 In the latter case, to which we refer as non-trivially integrable, it can be mapped to a system of  non-interacting fermions using the Jordan-Wigner transformation \cite{Lieb}.

 \section{Duality Relation}
 \label{sec:dualityRel}
The entire information on the spectrum of the time evolution operator $U_N(T)\teq U_N^T$, where \(U_N\) is the corresponding Floquet operator, is stored in the traces of its powers,
\begin{equation}
 Z(N,T)=\trace{U_N^T} \,. \label{trace}
\end{equation}
As a first step, we show that  $Z(N,T)$ can be equivalently represented  as  a partition function of a classical 2-dimensional Ising model defined on a $T\times N$ cyclic lattice.  By inserting the full set of states after each time step, we find
\begin{equation}
\fl
Z(N,T)=\sum_{\{\sigma_{n,t}= \pm 1\}}\langle\bm\sigma_1| \UIh \UKh |\bm\sigma_2\rangle\langle\bm\sigma_2| \UIh \UKh |\bm\sigma_3\rangle\cdots \langle\bm\sigma_T| \UIh \UKh  |\bm\sigma_1\rangle, \label{traces}
\end{equation}
where $|\bm\sigma_t\rangle=|\sigma_{1,t}\rangle \otimes|\sigma_{2,t}\rangle\otimes \dots \otimes|\sigma_{N,t}\rangle$, and  each $|\sigma_{n,t}\rangle$ is an  eigenstate of the spin operator $\hat{\sigma}^z_{n}$ at the $n$-th site with the eigenvalues $\sigma_{n,t}\teq\pm 1$. Since $|\bm\sigma_t\rangle$ are eigenstates of $\UIh$  we only need to study the matrix elements of $\UKh$ in more detail.

The kick operator factorizes into kick operations \(\UKh^{(i)}\) for the single spins \(i\) for which we can use the relation
\begin{equation}
\UKh^{(i)}=\eu^{-\iu\,{\bm b}\cdot \bm{\sigma}_i}=\cos{b} -\iu\left( \sin{\varphi}\, \sigma_i^x+\cos{\varphi}\, \sigma_i^z \right)\sin{b}
\,.
\end{equation}
This implies the following form for the different matrix elements,
\begin{eqnarray}
\label{matrixel}
\left\langle+1|\UKh^{(i)}|+1\right\rangle&=\cos b-\iu \cos\varphi\sin b={\rm e}^{-\iu K}{\rm e}^\eta{\rm e}^{-ih}\,, \nonumber
\\
\left\langle-1|\UKh^{(i)}|-1\right\rangle&=\cos b+\iu \cos\varphi\sin b={\rm e}^{-\iu K}{\rm e}^\eta{\rm e}^{ih}\,,
\\
\left\langle+1|\UKh^{(i)}|-1\right\rangle&=\left\langle-1|\UKh^{(i)}|+1\right\rangle=-\iu\sin\varphi\sin b={\rm e}^{\iu K}{\rm e}^\eta
\nonumber\,,
\end{eqnarray}
of the operator $\UKh$ in the $|\pm1\rangle$ basis.
The complex quantities $K$, $\eta$ and $h$ are given in terms of $b$ and $\varphi$ as 
\begin{equation}
{\rm e}^{-4iK}=1-\frac{1}{{x}^2}
\,,\quad
\eu^{4\eta}= x^2(x^2-1)
\,,\quad
\eu^{-2ih} =\frac{\cos b-i\sin b \cos \varphi}{\cos b+i\sin b \cos \varphi} 
\label{eq:khhparam}
\end{equation}
with $x\teq\sin{b}\,\sin{\varphi}$. The ansatz on the right side of (\ref{matrixel}) allows one to rewrite the matrix elements of $\UKh$ in the form
\begin{equation}
\fl
\langle\bm\sigma_t|  \UKh |\bm\sigma_{t+1}\rangle=\exp\left[-\iu \sum_{n=1}^N  \left(\frac{h}{2}( \sigma_{n,t}+\sigma_{n,t+1})+ K\sigma_{n,t}\sigma_{n,t+1} +\iu\eta\right)\right]\,.
\end{equation}
Including $\UIh$ we may therefore cast (\ref{traces})  into the form
\begin{equation}
\fl
 Z(N,T)=\sum_{\{\sigma_{n,t}= \pm 1\}}\exp\left(-\iu\sum_{n=1}^N\sum_{t=1}^T\left(
  J\sigma_{n,t}\sigma_{n+1,t}
 +K\sigma_{n,t}\sigma_{n,t+1}
 +h\sigma_{n,t}
 +i\eta\right)\right). \label{partfunction}
\end{equation}
Apart from the constant factor  $\eu^{NT\eta}$ the equation above is identical to the partition function of a 2-dimensional classical Ising model with complex coupling constants $J,K$ and $h$. Within this classical model, \(h\) plays the role of a magnetic field and the model is exactly solvable if $h$ vanishes. This corresponds to the non-trivially integrable regime of the KIC.

\begin{figure}
\centering
\includegraphics[width=0.4\textwidth]{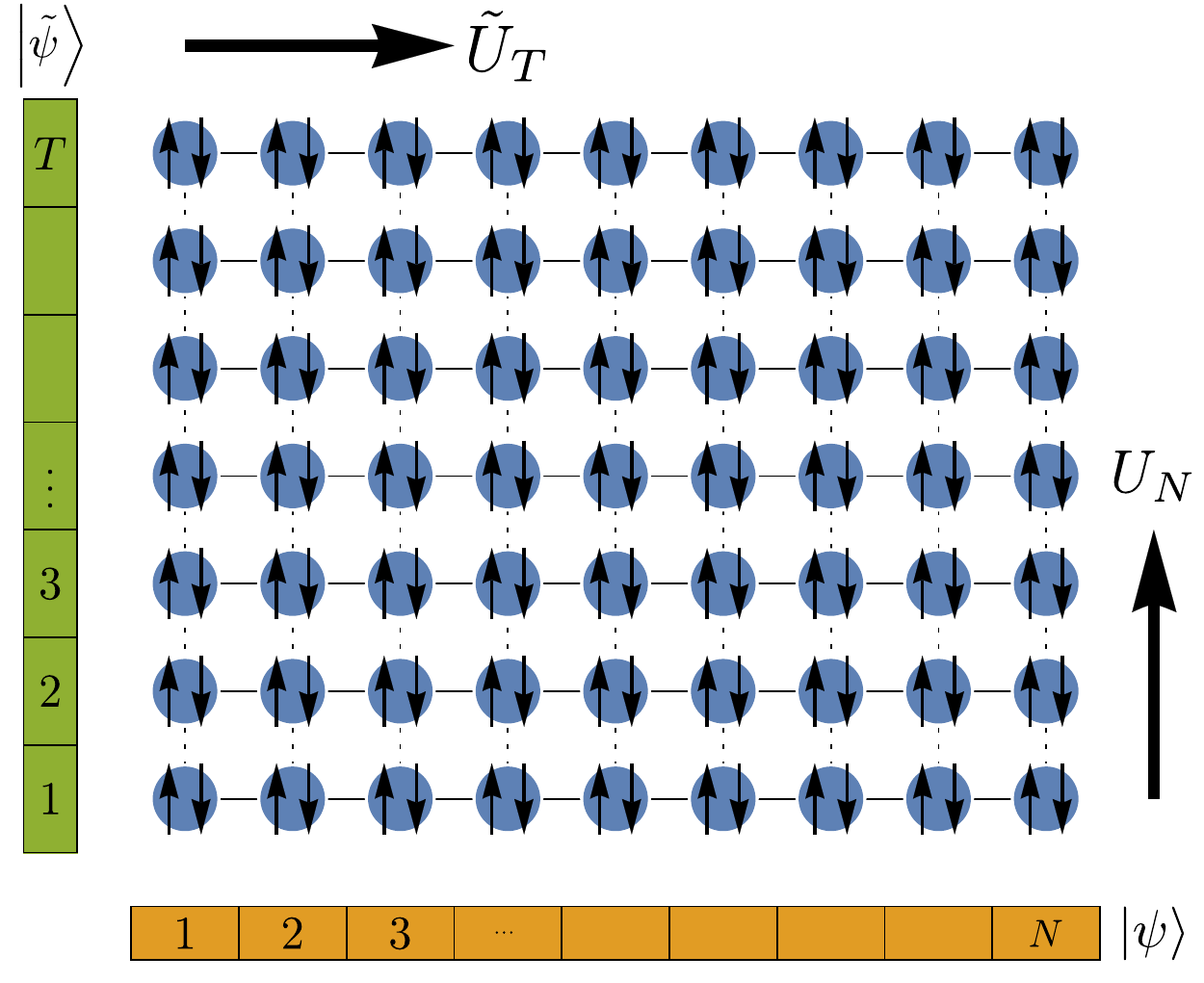}
\caption{A trace of the evolution operator \(U_N\), which evolves the usual \(N\)-particle states \(\left|\psi\right>\) in time, in the power \(T\) can be represented as a classical partition function of a periodic two dimensional \(N\times T\) Ising model. Instead of a sum over all quantum states the summation then contains all possible spin up or down configurations of the local sites. In the same fashion this sum can be contracted again onto a new operator \(\tilde{U}_N\) which effectively ``propagates'' states for \(T\) spins in particle direction.}
\label{fig:isingDiag}
\end{figure}

Equation (\ref{trace}) shows that the $2^N\times 2^N$ matrix  $U_N$ can be viewed as the transfer operator of the corresponding Ising model. The representation  (\ref{partfunction})  is symmetric under the exchange of time and particle 
indices $n\leftrightarrow t $, $N\leftrightarrow T$, if $J$ and $K$ are also exchanged.
Using this exchange symmetry, it is natural to consider the dual  transfer operator $\tilde{U}_T$ with a dimension of $2^T\!\times\! 2^T$ such that 
\begin{equation}\label{partfunc}
 Z(N,T)=\trace{\tilde{U}_T^N}\,.
\end{equation}
Whereas $U_N$ determines the time evolution of all $N$ particles by one time step, $\tilde{U}_T$ is valid for all times up to $T$ but only describes the relation  of neighboring spins, compare figure \ref{fig:isingDiag}.
This operator can again be split into Ising and kick parts,
\begin{equation}
 \tilde{U}_T=g^T\UI(K) \UK(\tilde{b},\tilde{\varphi})\,,\label{dualmatrix}
\end{equation}
where the dual parameters are determined  by
\begin{equation}\label{dualrela}
\qquad {\rm e}^{-4\iu J}=1-\frac{1}{\tilde{x}^2}, \quad \tilde{x}=\sin \tilde{b}\sin \tilde{\varphi}, \quad g^4=\frac{x^2(x^2-1)}{\tilde{x}^2(\tilde{x}^2-1)}.
\end{equation}
These equations are obtained by an ansatz of the same form as in (\ref{matrixel}), now for $\tilde{U}_T$ with the replacements $b\rightarrow\tilde{b}$, $\varphi\rightarrow\tilde{\varphi}$ and the exchange $J\leftrightarrow K$.
However, these new parameters are not real anymore causing \(\tilde{U}_T\) to be (generically) non-unitary. In \(\UK(\tilde{b},\tilde{\varphi})\) they give rise to a parameter \(\tilde{\eta}\) different from \(\eta\). We take this into account by introducing
\begin{equation}
g\teq\eu^{\eta-\tilde{\eta}}\,,
\end{equation} which yields the third of equations \eqref{dualrela}. 
The parameter $h$ in the dual picture remains unchanged as compared to the original \(U_N\). With (\ref{matrixel}) we find the following relation
\begin{equation}\label{hgleich}
 \tan b \cos \varphi=\tan \tilde b \cos\tilde\varphi
\end{equation}
between $b$, $\varphi$ and $\tilde{b}$, $\tilde{\varphi}$.
As a result of this change of viewpoint we have the following exact identity between traces of the unitary evolution for $N$-particles and the non-unitary ``evolution'' operator for $T$-particles
\begin{equation}
  \trace{{U}_N^T}= \trace{\tilde{U}_T}^N\,.\label{duality}
\end{equation}
The matrix \(U_N\) has dimension \(2^N\!\times\!2^N\).
This imposes severe limitations concerning the study of large particle numbers. In contrast, the dimension of \(\tilde{U}_T\) is independent of \(N\), but given by \(2^T\!\times\!2^T\).

\begin{figure}
\centering
\includegraphics[width=0.9\textwidth]{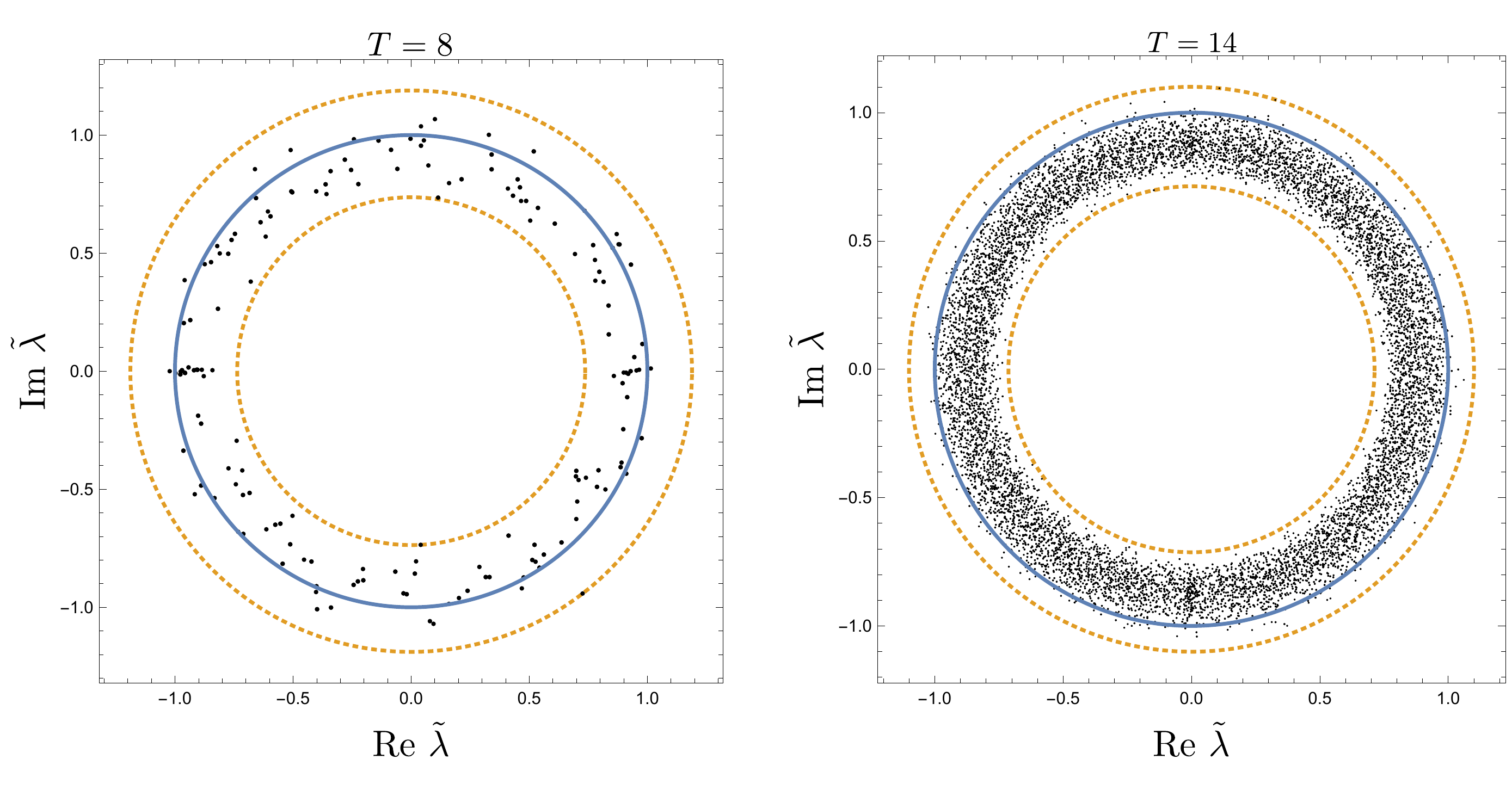}
\caption{Complex spectrum \(\{\tilde{\lambda}_i\}\) of the dual operator for $T=8$ (left) and $T=14$ (right). The parameters are identical in both cases and within the chaotic region: $J=0.7$, \(b\teq 0.9\sqrt{2}\), \(\varphi\teq \pi/4\). The blue (full) line is the unit circle, the red (dashed) lines indicate the edges of the spectrum.}
\label{fig1}
\end{figure}

\begin{figure}
\centering
\includegraphics[width=0.9\textwidth]{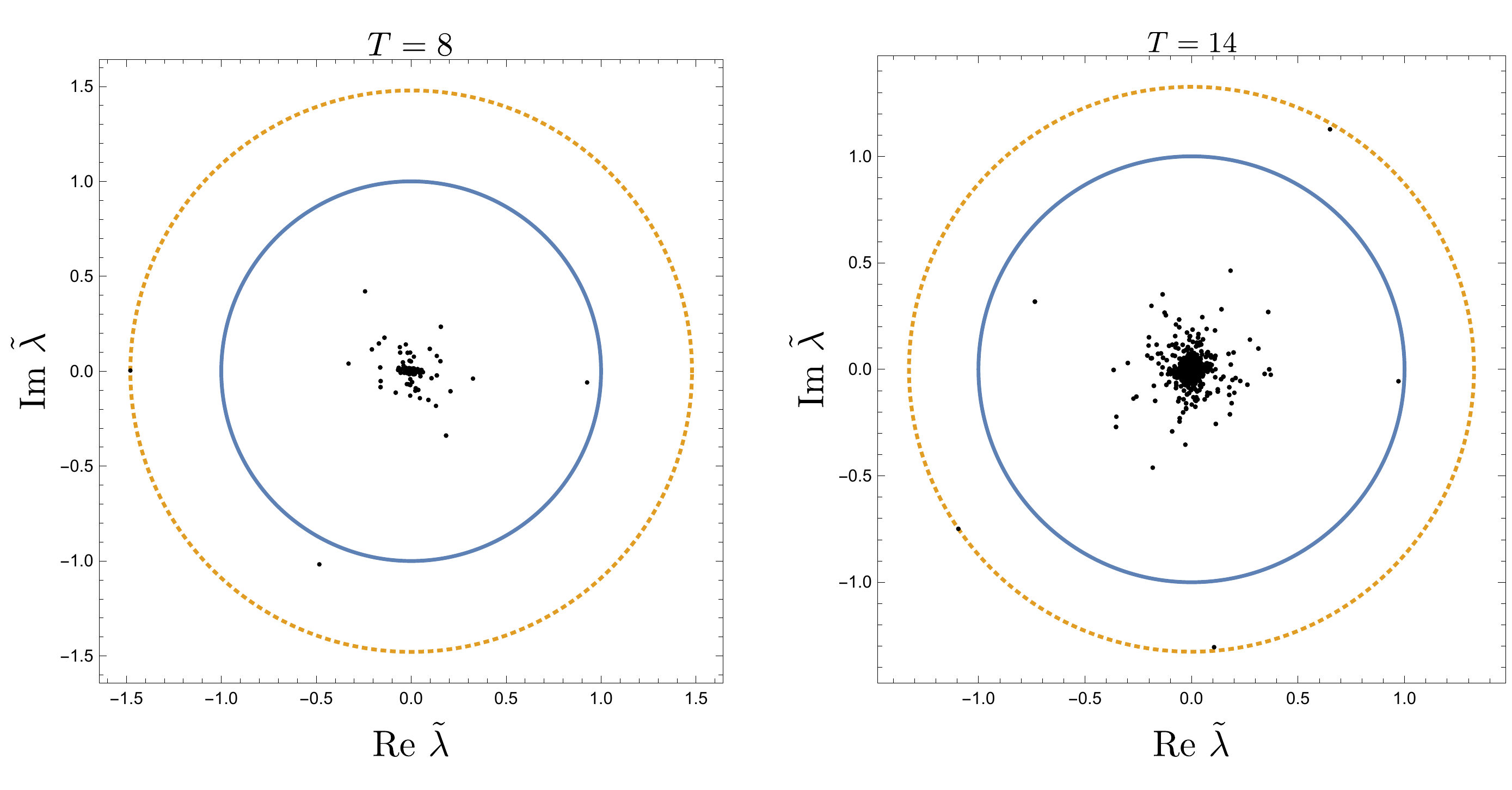}
\caption{Spectrum of the dual operator for $T=8$ (left) and $T=14$ (right). The parameters are identical in both cases and close to the trivially integrable region: $J=0.7$, \(b\teq 0.9\sqrt{2}\), \(\varphi\teq \pi/15\).}
\label{fig2}
\end{figure}

\begin{figure}
\centering
\includegraphics[width=0.9\textwidth]{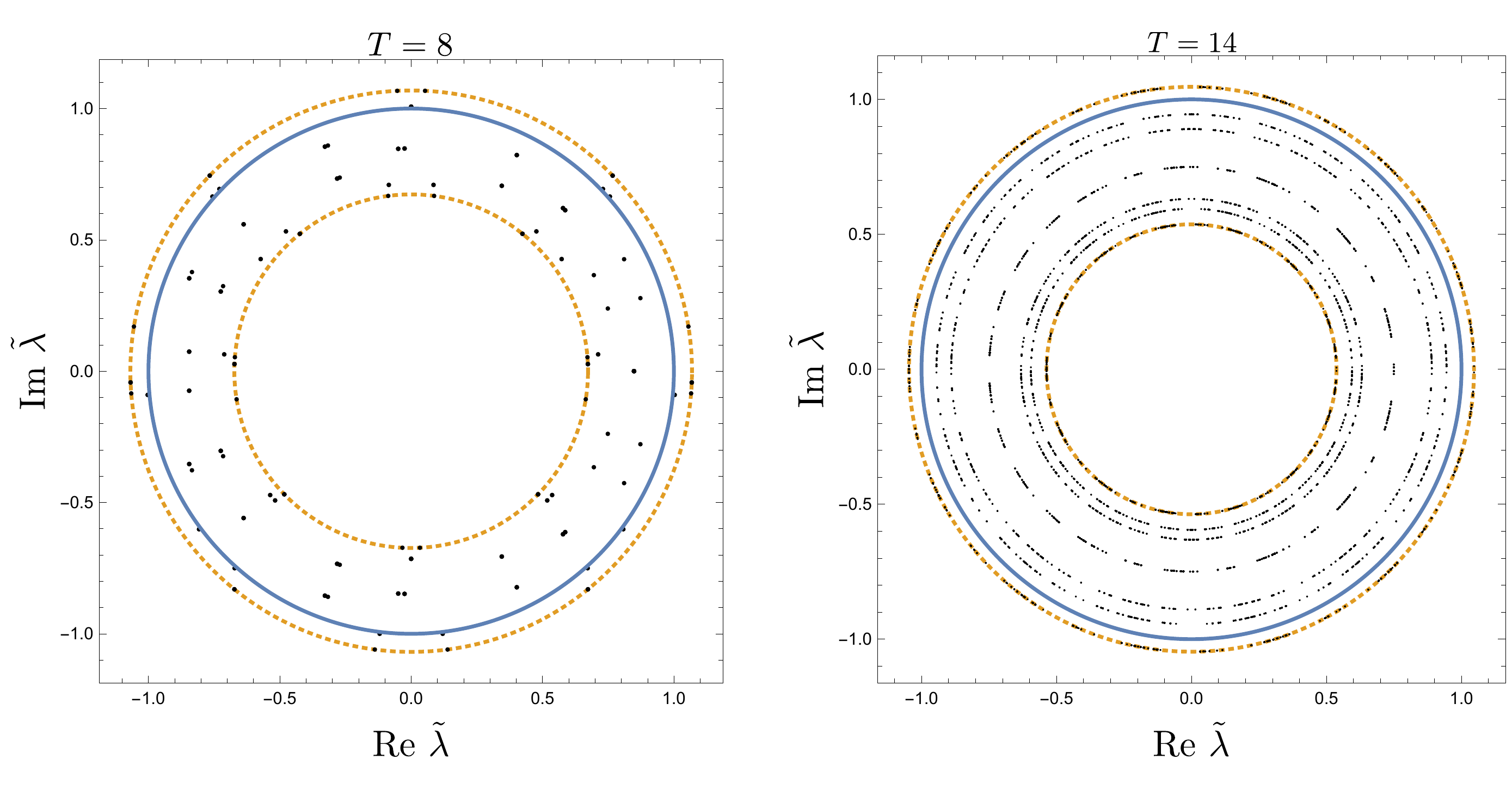} \\
\includegraphics[width=0.9\textwidth]{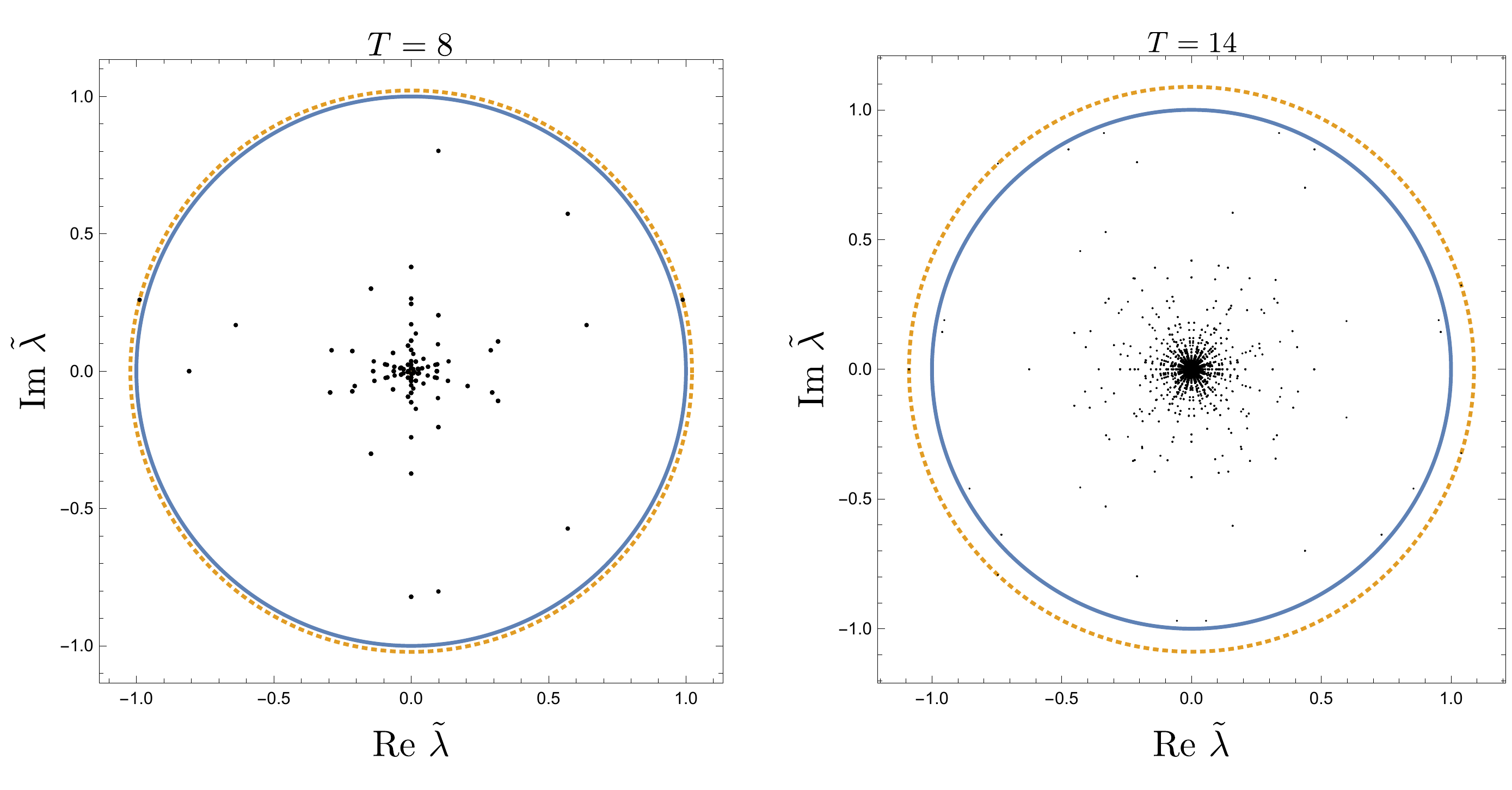} \\
\caption{At the non-trivially integrable point (\(\varphi\teq\pi/2\)) the behavior of the dual spectrum strongly depends on the choice of parameters. The upper row shows \(J\teq 0.7\) and \(b\teq 0.9\) (corresponding to \(n\teq 3\) (left) and \(n\teq 6\) (right)), while in the lower one \(b\) is changed to \(b\teq 0.9\sqrt{2}\) (\(n\teq 1\) resp. 2).}
\label{fig4}
\end{figure}

\section{Spectrum of the Dual Operator}
\label{sec:specDOp}
By the duality relation \eqref{duality} the entire information on the spectrum of $U_N$ can, in principle, be extracted from  the spectra of the dual operators $\tilde{U}_T$ for $T=1,2, \dots\,$. Thus,  it is   of  great interest to understand  how the spectrum of $ \tilde{U}_T$ depends on the system parameters and time $T$. 

We start by providing numerically calculated  eigenvalue dis\-tri\-bu\-tions of $ \tilde{U}_T$ that are typical  for the different parameter regimes of $J$ and $\bm b$.
Afterwards, we explain in detail how the spectrum of the dual operator is obtained from  (\ref{partfunc})-(\ref{hgleich}), considering the non-trivially integrable case.

The properties  of the dual eigenvalue spectrum \(\{\tilde{\lambda}_i\}\) crucially depend  on the regime considered.
In the chaotic regime, see figure \ref{fig1}, the bulk of the eigenvalues is placed within the unit disc, 
such that its boundary serves as the edge of the spectrum. Increasing $T$ this boundary becomes sharper while the  gap between the largest eigenvalues and the bulk of the spectrum shrinks to zero. Generically, \(\tilde{U}_T\) is non unitary. However, along the line \(J\teq\pi/4,\,\varphi\teq\arcsin{(\sqrt{2}\sin{b})^{-1}}\) and \(b\in[\pi/4,\pi/2]\) we find \(J\teq K,\,b\teq \tilde{b},\,\varphi\teq\tilde{\varphi}\) which implies that both \(U_N\) and \(\tilde{U}_T\) are unitary and differ only by their dimension. This special case is reminiscent of \cite{Gutkin}, where the dual operator is strictly unitary.

Close to the trivially integrable regime, as seen in figure \ref{fig2}, most of the eigenvalues are localized in the vicinity of zero with a large gap separating the bulk from the largest eigenvalues. In the limit of $b^x\to 0$ the entire  spectrum except for two eigenvalues collapses  to zero.
In the non-trivially integrable  regime $b^z\to 0$ the spectrum of  $ \tilde{U}_T$ has  a regular structure, see figure \ref{fig4}, with large degeneracies in the  absolute values of the eigenvalues.
In this case we can distinguish two qualitatively different regimes, the upper row in figure \ref{fig4} is reminiscent of the chaotic regime with its minimal inner radius. In the lower one, the inner gap is closed and this reminds visually of the trivially integrable regime.
Understanding the distribution of the eigenvalues in more detail is possible, as they are analytically accessible by a map on an equivalent  system of free fermions.

For the sake of simplicity, we only consider spin-chains with an even number \(N\) of total spins. But it is straightforward to generalize this to odd $N$, see \ref{appendA} for details.
If the parameters of the operator \(U_N\) belong to the non-trivially integrable regime this holds as well for the dual operator as can be inferred from (\ref{hgleich}) because \(\cos{\varphi}\teq 0\) implies \(\cos{\tilde{\varphi}}\teq 0\). In this case both of the other parameters, \(K\) and \(\tilde{b}\), turn out to be imaginary up to a constant real part,
\begin{eqnarray}
\label{eq:NTDualParams}
K& = -\frac{\pi}{4}+\frac{\iu}{4} \ln{\cot^2{b^x}}\,,
\\ \nonumber
\tilde{b}& =  \text{arccot}\exp{\left(-\iu \frac{\pi}{2}-2\iu J\right)}=
-\frac{\pi}{4}+\frac{\iu}{4} \log{\cot^2{J}}\,.
\end{eqnarray}
In \ref{appendA} we derive an analytic form of the eigenvalues of \(U_N\) and \(\tilde{U}_T\).
In both cases the eigenvalues are given by structurally similar  combinatorial products.  In particular,  for even $T$ each eigenvalue of  \(\tilde{U}_T\) can be labeled by a sequence of symbols $\bm\varepsilon=\varepsilon_1 \dots \varepsilon_T$,   $\varepsilon_j\in\{ 1,\dots, 4\}$  such that 
\begin{equation}
\tilde{\lambda}_{\bm \varepsilon} = \prod^{T/2}_{j=1} \Lambda_{\varepsilon_j}(k_j),  \qquad k_j ={\pi(2j+1)}/{T}.
\label{eq:lambdaComb:body}
\end{equation}
Here, each factor 
 \(\Lambda_{\varepsilon_j}(k_j)\) is drawn  out of a set of four elements,
\begin{equation}
\Lambda(k)= \{\mu_-(k),\,1, \,1,\,\mu_+(k) \}\,,
\label{eq:lambdaSet:body}
\end{equation}
where  \(\mu_+(k)\mu_-(k)=1\),   \(\mu_\pm(k)\teq \alpha(k)\pm \sqrt{\beta(k)}\), with real valued functions \(\alpha,\,\beta\) provided  in \ref{appendA}. The index \(\varepsilon_j\) denotes which of the elements contributes to \eqref{eq:lambdaComb:body}.

Whereas for \(U_N\) the absolute values of \(\mu_\pm(k)\) are  always  one,  making its spectrum unitary, we find that for \(\tilde{U}_T\) this is not necessarily the case. Instead, it depends on the sign of \(\beta(k)\). Specifically,  \(|\mu_\pm(k)|teq 1\) only if  \(\beta(k)\) is negative, in which case   $\mu_\pm(k)$ is complex, in the other case $\mu_\pm(k)$ is real.
It is therefore apparent that different \(\bm\varepsilon\) combinations which have identical values \(\varepsilon_j\) for all real \(\mu_\pm(k)\) lead to eigenvalues with identical absolute value. However, they must not necessarily ber degenerate as they can differ in phase due to the complex \(\mu_\pm(k)\).
Under the  variation of the system parameters it might happen that one of the \(\beta(k)\)'s changes its sign. At these parameters  a pair of  circles   merges  which  also leads  to a  change in the degeneracies of the absolute values of $\tilde{\lambda}_{\bm \varepsilon}$.

\begin{figure}
\centering
\includegraphics[width=0.4\textwidth]{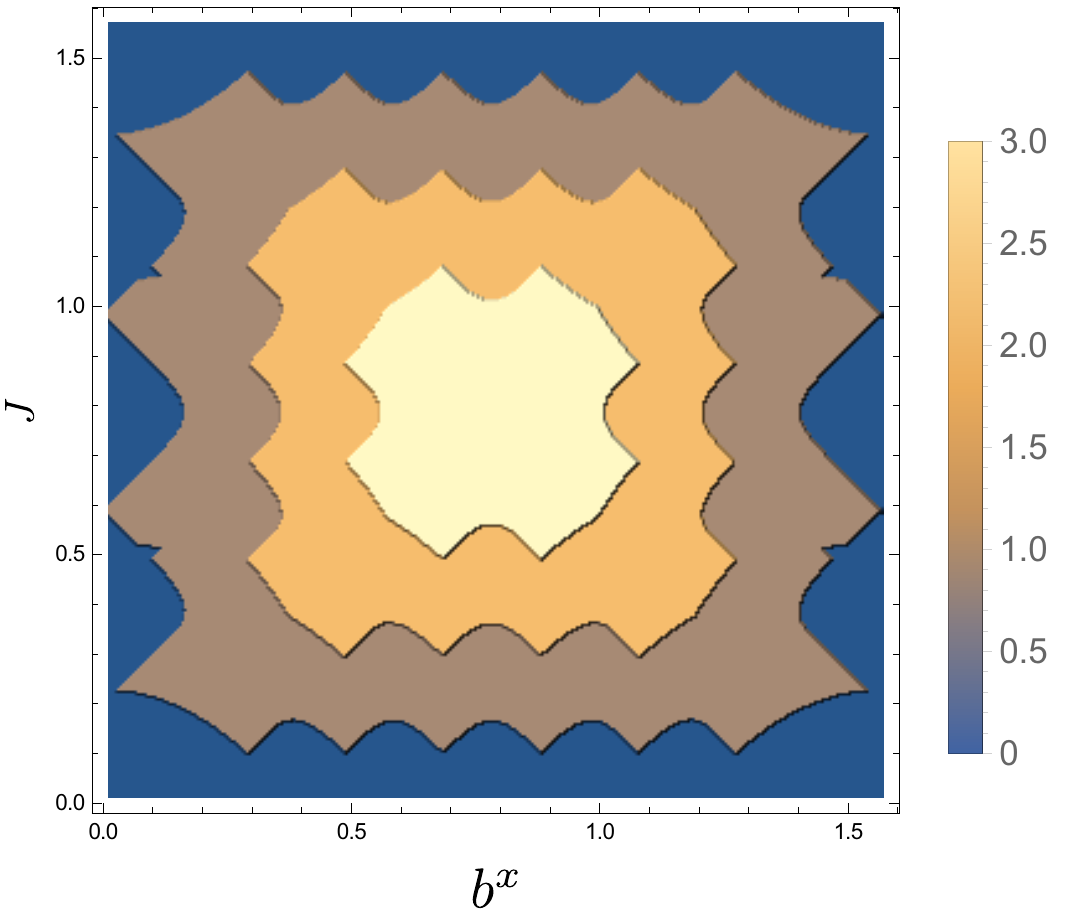}
\hspace{0.75cm}
\includegraphics[width=0.4\textwidth]{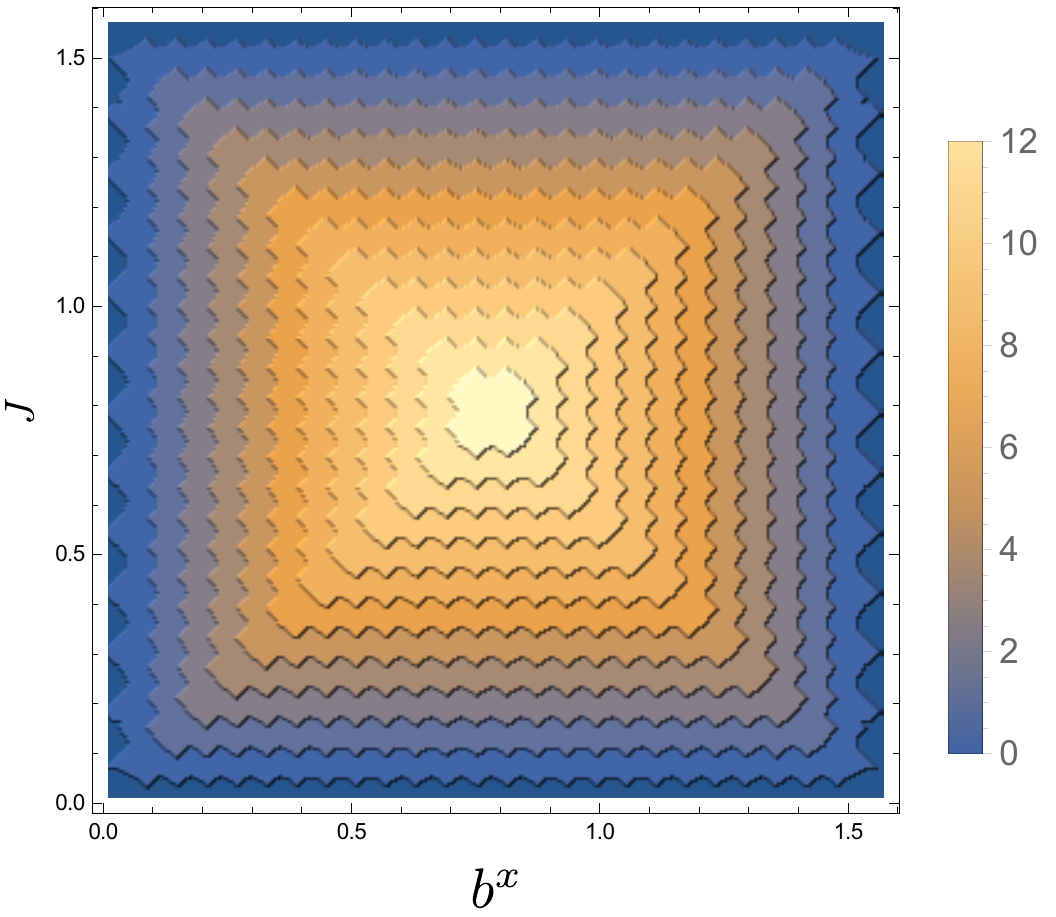}
\caption{Value of \(n\) for different parameters of \(J\) and \(b^x\) with either \(T\teq 8\) (left) or \(T\teq 26\) spins (right). The number of layers grows with \(T\) but retains the pyramidal form. One can notice a slight deviation from the mirror symmetry in the indention pattern for \(n\teq 0,\,1\), depending on whether one is close to the \(J\) or the \(b^x\) axes. Due to coloring this effect is barely visible on the right hand side. Along the diagonal \(J\teq b^x\) it can occur that both parities \(\mathcal{P}\) contribute to the largest eigenvalues, which is not resolved in the figure.}
\label{fig:pyramid}
\end{figure}

We  are particularly   interested  in  the  outer   circle of the spectrum,
\ie the  eigenvalues of \(\tilde{U}_T\)
with the largest magnitude,  as they provide the dominant contribution to traces of the evolution operator (see \eqref{duality}) in the large $N$ limit. The degeneracy of the eigenvalues is characterized by a non-negative integer parameter \(n \leq N/2 \) counting the number of sets \(\Lambda(k)\) with a negative \(\beta(k)\).  Figure \ref{fig:pyramid} shows its values in dependence of \(J\) and \(b\). For the  eigenvalues with the largest modulus  all \(\mu_\pm(k)\) for which \(\beta(k)>0\) have  to contribute in \eqref{eq:lambdaComb:body} leaving only \(n\) sets from which the entries can be chosen freely. As the entry $1$  appears twice in the set (\ref{eq:lambdaSet:body}) both choices do not affect the value of \(\tilde{\lambda}_{\bm \varepsilon} \) and thus lead to a degeneracy. 

In general, the spectrum at each  circle can be split into $\ell$ multiplets, where the $i$-th multiplet \(i=1, \dots \ell \) is composed of   \(m_i \) distinct eigenvalues  having the same degeneracy \(d_i\). By simple combinatorial arguments  we show in  \ref{appendA} that  for  the outer spectral circle  these numbers are given by
\begin{equation}
d_i=2^{2i},
\qquad
m_i=2^{n-2i}\binom{n}{2i}, \quad \ell =\lfloor n/2 \rfloor\teq \operatorname{floor}{(n/2)}
\,.
\label{eq:degDef}
\end{equation}
Using \eqref{eq:degDef} we can express the total number of eigen\-values  as
\begin{equation}
\sum_{i=0}^{\lfloor n/2 \rfloor} d_i m_i = 2^{2n-1}
\label{sumTotalEV}
\end{equation}
and the number of distinct eigenvalues by
\begin{equation}
\sum_{i=0}^{\lfloor n/2 \rfloor} m_i=(1+3^n)/2\,.
\label{sumDistinctEV}
\end{equation}
As figure \ref{fig:pyramid} suggests we find abrupt transitions of these numbers in the parameter space which occur whenever the outer circle is replaced by one of the inner circles taking on its role.

\section{Conclusion}\label{section5}
For \(N\)-particle systems the Hilbert space dimension grows exponentially with \(N\) which makes it challenging to consider systems built up by more than a few constituents, even if limited to nearest neighbor interaction as in the considered model.
The duality approach presented here allows the exact calculation of traces of the time evolution operator \(U_N\) in power \(T\) for arbitrary particle numbers but short times \(T\). This gives access to a regime opposite to the usually considered limits of few particles but large times. To achieve this we relate the trace of \(U_N^T\) to the trace of a dual operator \(\tilde{U}_T^N\) whose dimension \(2^T\!\times\! 2^T\) is in this regime significantly smaller than the dimension \(2^N\!\times\! 2^N\) of the original matrix.
Up to a multiplicative correction factor this dual operator is still described by an evolution operator for the KIC, albeit (typically) for complex parameters.
Our approach indicates that for short times the system can not explore the full Hilbert space due to the short ranged interaction, in a broad sense such a reduction in complexity can also be used for the propagation of states and the calculation of their expectation values \cite{entangle1,entangle2}.

Due to these complex parameters, it turns out that \(\tilde{U}_T\) is non-unitary and features a rich structure, which depends on the parameter regime considered.
For instance, we find a sharp contrast between the chaotic regimes, where  the eigenvalues of \(\tilde{U}_T\) posses an inner gap, and the trivially integrable regime, where it is not present.
In the chaotic case the eigenvalues are distributed over a ring, while in the integrable regimes they are highly ordered. In the non trivially regime they are situated on concentric circles and highly degenerate, close to the trivially integrable regime all besides two eigenvalues tend to zero.
Given that the KIC model does not have a classical limit,
 chaos in this system can not be defined via the dynamics of an underlying classical model. Instead one has to refer to spectral statistics, for instance to the level spacing, or to correlators of the evolution operator \(U_N\). The structural properties of \(\tilde{U}_T\) might yield an additional approach to analyze such properties and transitions between the regimes.

Especially in the intermediate regimes the non-unitarity allows us to approximate powers of \(\tilde{U}_T\) by the largest eigenvalues, a property which we use heavily to determine asymptotic properties of the form-factor and the spectral density in the accompanying paper. In the chaotic case no longer single eigenvalues are important but instead the distribution on the outer edge. It is an interesting question to which extend such a distribution can be modeled by a RMT ensemble of non-unitary matrices. This could give further insight into the originally posed question of universality for short times but large numbers of particles.

Finally, although we present the concept of the dual operator for the KIC only, it can be extended to a broader class of systems. For kicked systems with nearest neighbor interaction this is a straight forward step and may also include disorder in time or particle direction. Moreover, it is possible to consider interactions with \eg the second next spins, changing the constraint of nearest neighbor interaction into one of short range interaction.


\appendix
\section{Jordan Wigner Transform for the KIC}\label{appendA}
The Hamiltonian can be expressed after a Jordan Wigner transformation of the form
\begin{eqnarray}
\hatN{a}_n=\frac{1}{2}\left(\prod_{j=1}^{n-1}\hatN{\sigma}_j^x\right)\left(\hatN{\sigma}_n^z-i\hatN{\sigma}_n^y\right)\,,
\\
\hatN{a}_n^\dagger=\frac{1}{2} \left(\prod_{j=1}^{n-1}\hatN{\sigma}_j^x\right) \left(\hatN{\sigma}_n^z+i\hatN{\sigma}_n^y\right)
\,,
\\ 
\hatN{\sigma}_n^x=1-2\hatN{a}_n^\dagger \hatN{a}_n \,,
\end{eqnarray}
and the transformation to the Fourier domain 
\begin{equation}
 \hatN{b}_k=\frac{1}{\sqrt{N}}\sum_{n=1}^N{\rm e}^{ikn}\hatN{a}_n
\end{equation}
as
\begin{equation}
 H_I=2J\sum_k\left[\cos k\left(\hatN{b}_k^\dagger\hatN{b}_k-1/2\right)+\frac{i}{2}\left(\hatN{b}_k^\dagger \hatN{b}_{-k}^\dagger-\hatN{b}_{-k}\hatN{b}_k\right)\sin k\right]
\end{equation}
and 
\begin{equation}
 H_K=-2b\sum_{k}\left(\hatN{b}_k^\dagger \hatN{b}_k-1/2\right)\,.
\end{equation}
This transformation can be performed in the same way as in \cite{Lieb}. The allowed $k$-values depend
on the parity $\mathcal{P}=\prod_{i=1}^N\hatN{\sigma}_n^x=(-1)^\mathcal{N}$ with the particle number $\mathcal{N}=\sum_{n=1}^N\hatN{a}_n^\dagger \hatN{a}_n
=\sum_{k}\hatN{b}_k^\dagger \hatN{b}_k$ and are given by 
\begin{equation}
k_j=\frac{\pi}{N}
\left\{\begin{array}{lr}
2j & \mathcal{P}=-1\\
2j+1 & \mathcal{P}=+1
\end{array}
\right.\,,
\qquad
j=0,\,1,\,\ldots,\,N-1
\,.
\label{eq:kDef}
\end{equation}
Using the transformation 
\begin{equation}
 \hatN{\eta}_k=\cos\vartheta_k\hatN{b}_k-i\sin\vartheta_k \hatN{b}_{-k}^\dagger,\hspace*{2cm}\vartheta_k=k/2+\pi/2
\end{equation}
the Hamiltonian can be expressed as 
\begin{equation}
 \HI=-2J\sum_{k}\left(\hatN{\eta}_k^\dagger\hatN{\eta}_k-1/2\right)
\end{equation}
and 
\begin{equation}
\fl
\HK=-2b\sum_{k} \left[\cos2\vartheta_k \left(\hatN{\eta}_k^\dagger \hatN{\eta}_k-1/2
 \right)+\frac{i}{2} \sin2\vartheta_k \left(\hatN{\eta}_k^\dagger \hatN{\eta}_{-k}^\dagger -\hatN{\eta}_{-k} \hatN{\eta}_k\right)\right]\,.
\end{equation}
Using that $\hatN{b}_k$ and $\hatN{\eta}_k$ are fermionic operators we obtain for the Ising and kick part of 
the Floquet operator
\begin{eqnarray}
 U_I&=&{\rm e}^{2 i J\sum_{k}\left(\hatN{\eta}_k^\dagger\hatN{\eta}_k-1/2\right)}={\rm e}^{-iNJ}\prod_{k}\left[1+\left({\rm e}^{2iJ}-1\right)\hatN{\eta}_k^\dagger\hatN{\eta}_k\right]\nonumber\\
 U_K&=&{\rm e}^{2ib\sum_{k}\left(\hatN{b}_k^\dagger \hatN{b}_k-1/2\right)}={\rm e}^{-iN b}\prod_{k}\left[1+\left({\rm e}^{2ib}-1\right)\hatN{b}_k^\dagger \hatN{b}_k\right]\nonumber\\&=&
 {\rm e}^{-iNb}\prod_{k}\left[1+\left({\rm e}^{2ib}-1\right)\left(\cos^2\vartheta_k\hatN{\eta}_k^\dagger\hatN{\eta}_k+\sin^2\vartheta_k\hatN{\eta}_{-k}\hatN{\eta}_{-k}^\dagger\right.\right.\nonumber\\&&
 +\left.\left.\frac{i}{2}\sin2\vartheta_k
 \left(\hatN{\eta}_k^\dagger\hatN{\eta}_{-k}^\dagger-\hatN{\eta}_{-k}\hatN{\eta}_k\right)\right)\right]
\end{eqnarray}
In the latter expression $k$ is only coupled to itself and to $-k$, the Floquet operator thus splits into $4\!\times\! 4$ subblocks ($k$ and $-k$ occupied, $k$ and $-k$ unoccupied, only $k$ occupied, only $-k$ occupied) that can be diagonalized analytically. The resulting eigenvalues are the entries of the sets
\begin{equation}
\Lambda(k)= \{\mu_-(k),\,1, \,1,\,\mu_+(k) \}\,,
\label{eq:NTLambdaSet}
\end{equation}
where \(\mu_\pm(k)\teq \alpha(k) \pm \sqrt{\beta(k)}\) with
\begin{eqnarray}
\label{eq:AlphaBeta}
\fl
\alpha(k) = \frac{1}{4}{\eu}^{2 \iu \left(J+b^x\right)}\left[\left(1+\cos2\vartheta_k\right)\left(1+{\eu}^{-4\iu\left(J+b^x\right)}\right)
\!+\!
\left(1-\cos2\vartheta_k\right)
 \left(\eu^{-4\iu J}+\eu^{-4\iu b^x}\right)\right]\,,
\\ \fl 
\beta(k)=  \frac{\eu^{4\iu (J+b^x)}}{16}\left(\left(1+\eu^{-4\iu J}\right)\left(1+\eu^{-4\iu b^x}\right)+\left(1-\eu^{-4\iu J}\right)\left(1-\eu^{-4\iu b^x}\right)\cos2\vartheta_k
 \right)^2-1\,.\nonumber
\end{eqnarray}
These quantities are symmetric under the exchange of \(J\) and \(b^x\) making it an almost exact symmetry of the non-trivially integrable system, a small deviation is discussed below.
Furthermore, they fulfill the relation
\begin{equation}
\label{eq:LambdaRel}
\mu_+(k) \mu_-(k) =1\,.
\end{equation}
The corresponding occupation numbers to \(\Lambda(k)\) in \eqref{eq:NTLambdaSet} are \(\{0,1,1,2\}\).

Further on, we obtain two special sectors for \(k\teq 0\,,\pi\) which are not paired to any \(-k\) sectors. The eigenvalues for those cases are
\begin{equation}
\begin{array}{ll}
k=0: & \Lambda(0)=\{\eu^{\iu(J-b^x)},\,\eu^{\iu(b^x-J)} \} \\
k=\pi: & \Lambda(\pi)=\{\eu^{-\iu(J+b^x)},\,\eu^{+\iu(J+b^x)} \} \\
\end{array}
\label{eq:specialK}
\end{equation}
for an occupation value of \(0\) (left eigenvalue) and single fermion occupation (right), respectively.
In the \(\hatN{\eta}\)-basis the total occupation number \(\sum_{k}\hatN{\eta}_k^\dagger \hatN{\eta}_k\) has to be even for both parities. Apart from this constraint the eigenvalues of \(\hatN{U}_N\) are combinatorial products,
\begin{equation}
\lambda = \prod_{i} \Lambda_{\sigma(i)}(k_i)\,,
\label{eq:NTCombProd}
\end{equation}
whose components \(\Lambda_{\sigma(i)}(k_i)\) are chosen from the sets \eqref{eq:NTLambdaSet} and \eqref{eq:specialK}.

To extend this result to the dual picture, \ie the eigenvalues of \(g^{-T} \tilde{U}_T\), it is necessary to replace \(J\) by \(K\), \(b^x\) by \(\tilde{b}\), as given by \eqref{eq:NTDualParams}, and \(N\) by \(T\) in the definition of \(k_j\) in \eqref{eq:kDef}. This leads to purely real, but not necessarily positive, functions \(\alpha(k)\) and \(\beta(k)\). 
Due to \eqref{eq:LambdaRel}, in the case where \(\beta(k)\) is negative, \(\mu_\pm(k)\) are complex conjugated numbers with absolute value 1 while in the case \(\beta(k)>0\) they are real and the absolute value of one of them is larger than unity.
Throughout the remaining discussion \(n\leq N/2\) denotes the number of negative radicands of \(\sqrt{\beta(k)}\).

To simplify the discussion we restrict ourselves, at first, to the case of \(N\) even and \(\mathcal{P}\teq +1\), where none of the special sectors enter \eqref{eq:NTCombProd}. As pointed out before, we are interested in the largest eigenvalues.
In  these cases the combinatorial products must contain all \((N/2-n)\) of the \(\mu_\pm(k)\) with \(|\mu_\pm(k)|>1\). All of those entries belong to an even occupation number and possess a \(\beta(k)>0\).
The remaining \(n\) choices for the other factors \(\Lambda_{\sigma(i)}(k_i)\) only have an influence on the phase, not on the absolute value. We therefore have freedom in combining them. In general, when all \(\mu_\pm(k)\) are different, we can choose
all \(2^n\) possible combinations, each yielding a different phase and therefore eigenvalues of the same magnitude. Next, we are allowed to replace two (to keep an even occupation number) of the complex \(\mu_\pm(k)\) by one of the two unit elements within the respective \(\Lambda(k)\) set. This eigenvalue is therefore \(2^2\)-fold degenerate and we find \(2^{n-2} \binom{n}{2} \) different eigenvalues of this type. Replacing more $\mu_\pm(k)$ with absolute value one by unit elements, the degeneracies grow in multiples of 4 up to a maximum of \(2^{2m}\) with \(2m\teq n\) for even \(n\) (\(2m\teq n-1\) if odd), which occurs if all ``complex'' contributions to the eigenvalue are completely replaced by ones. Such an eigenvalue can only occur once (\(2\binom{n}{2m}\teq 2 n\) times for odd \(n\), as one complex \(\mu_\pm\) is left).
This leads us to the degeneracies \(d_i\) and multiplicities \(m_i\) in \eqref{eq:degDef}.

In case of the other parity (\(\mathcal{P}\teq -1\)) the picture looks slightly more complicated, as now the special sectors in \eqref{eq:specialK} have to be taken into account.
The maximal absolute value resulting from these sets in the dual picture is \(\eu^{2 |\Im \tilde{b} |}\) for an even and  \(\eu^{2 |\Im K |}\) for an odd occupation. Assuming the even contribution is the larger one we can select the remaining \( \Lambda(k) \) as discussed before. However, in the odd case the larger of either \(\eu^{2 |\Im \tilde{b} |}\) together with all \(\mu_\pm(k)\)  or \(\eu^{2 |\Im K |}\) together with all but the smallest \(\mu_\pm(k)\) (which has an absolute value larger 1 if \(n \teq 0\)) has to be chosen. In the latter case the equations for \(d_i\) and \(m_i\) are slightly modified by replacing \(2 i \to 2i+1\) with \(i\teq 0,\,1,\,\ldots,\lfloor (n-1)/2 \rfloor \) (in the special case of \(n\teq 0\) this implies only one doubly degenerate eigenvalue). This modification leaves the total number of eigenvalues \eqref{sumTotalEV} invariant, however the distinct eigenvalues \eqref{sumDistinctEV} are then given by \((3^n -1)/2\).


\section*{References}
\begin{thebibliography}{10}
\bibitem{haake}S.\ M\"uller, S.\ Heusler, P.\ Braun, F.\ Haake, A.\ Altland,
Phys.\ Rev.\ Lett.\ {\bf93}, 014103 (2004).
\bibitem{haake2} S.\ Heusler, S.\ M\"uller, A.\ Altland, P.\ Braun, F.\ Haake, Phys.\ Rev.\ Lett.\ {\bf98}, 044103 (2007).
\bibitem{haake3} F.\ Haake, \textit{Quantum Signature of Chaos}, Springer 2010.
\bibitem{stoeckmann} H.J.\ St\"ockmann, \textit{Quantum Chaos -- an introduction}, Cambridge University Press (2006).
\bibitem{Berry} M.V.\ Berry, Proc.\ R.\ Soc.\ A {\bf400}, 229 (1985).
\bibitem{regen1} T.\ Engl, J.\ Dujardin, A.\ Arg\"uelles, P.\ Schlagheck, K.\ Richter, J.D.\ Urbina,
Phys. Rev. Lett. {\bf 112}, 140403 (2014).
\bibitem{haemmer}  J.\ H\"ammerling, B.\ Gutkin, T.\ Guhr,
EPL {\bf 96}, 20007 (2011).
\bibitem{haemmer2}  J.\ H\"ammerling, B.\ Gutkin, T.\ Guhr
J. Phys. A {\bf43}, 265101 (2010).
\bibitem{dubert} R.\ Dubertrand, S.\ M\"uller, \texttt{arXiv:1509.02339}.
\bibitem{buchleit1} M.\ Gessner, V.M.\ Bastidas, T.\ Brandes, A.\ Buchleitner, \texttt{arXiv:1509.08429}.
\bibitem{znidaric} M.\ \v{Z}nidari\v{c}, Phys.\ Rev.\ Lett. {\bf112}, 040602 (2014).
\bibitem{buca} B.\ Bu\v{c}a, T.\ Prosen, Phys.\ Rev.\ Lett. {\bf112}, 067201 (2014).
\bibitem{Gutkin} B.\ Gutkin, V.\ Osipov, Nonlinearity {\bf29}, 325 (2016).
\bibitem{prosenJt-2}
T.\ Prosen, Prog. Theor. Phys. Suppl. {\bf139}, 
191 (2000).
\bibitem{prosen2}
T.\ Prosen,
Phys.\ Rev.\ E {\bf65},
036208 (2002).
\bibitem{prosen2007}
C.\ Pineda, T.\ Prosen,
Phys.\ Rev.\ E {\bf76},
061127 (2007).
\bibitem{prosenB3-d}
T.\ Prosen, Journal of Phys.\ A {\bf40}, 7881 (2007).
\bibitem{Lieb} E.\ Lieb, T.\ Schultz, D.\ Mattis, Ann.\ of Phys.\ {\bf16}, 407 (1961).
\bibitem{entangle1} M.C.\ Ba\~nuls, M.B.\ Hastings, F.\ Verstraete, J.I.\ Cirac, Phys.\ Rev.\ Lett.\ {\bf102}. 240603 (2009).
\bibitem{entangle2} M.B.\ Hastings, R.\ Mahajan, Phys.\ Rev.\ A\ {\bf91}, 032306 (2015).
\end{thebibliography}
 \end{document}